# USING ENSEMBLE LEARNING WITH HYBRID GRAPH NEURAL NETWORKS AND TRANSFORMERS TO PREDICT TRAFFIC IN CITIES

**ISMAIL ZRIGUI[1], SAMIRA KHOULJI[1], MOHAMED LARBI KERKEB**

1 Innovate Systems Engineering Lab (ISI), National School of Applied Sciences, Abdelmalek Essaadi University, Tetouan, Morocco

E-mail : izrigui@uae.ac.ma

**ABSTRACT**

Intelligent transportation systems (ITS) still have a hard time accurately predicting traffic in cities, especially in big, multimodal settings with complicated spatiotemporal dynamics. This paper presents HybridST, a hybrid architecture that integrates Graph Neural Networks (GNNs), multi-head temporal Transformers, and supervised ensemble learning methods (XGBoost or Random Forest) to collectively capture spatial dependencies, long-range temporal patterns, and exogenous signals, including weather, calendar, or control states. We test our model on the METR-LA, PEMS-BAY, and Seattle Loop tree public benchmark datasets. These datasets include situations ranging from freeway sensor networks to vehicle-infrastructure cooperative perception. Experimental results show that HybridST consistently beats classical baselines (LSTM, GCN, DCRNN, PDFormer) on important metrics like MAE and RMSE, while still being very scalable and easy to understand. The proposed framework presents a promising avenue for real-time urban mobility planning, energy optimization, and congestion alleviation strategies, especially within the framework of smart cities and significant events such as the 2030 FIFA World Cup.

## 1  Introduction

Major cities' mobility patterns have been significantly changed by urban expansion and the ongoing increase in the usage of private vehicles. Persistent issues brought about by this quick change include increased greenhouse gas emissions, recurrent traffic, and a continuous drop in the standard of living for locals [1], [2]. Transportation-related activities alone account for about 80% of total $CO_2$ emissions in a number of urban areas [3]. Because of this, the transportation sector has become a crucial area for smart mobility and sustainable development projects. Intelligent transportation systems (ITS) are now one of the best ways to solve urban traffic problems since they make networks perform better, cut down on travel delays, and have less of an influence on the environment [4].

Early work on traffic forecasting mainly relied on linear statistical frameworks such as Kalman filters and ARIMA.Whereas classical models could reveal some periodic pattern, yet their performance degraded rapidly when confronted with the complex, irregular, and nonlinear behavior characteristic of real traffic systems [5]. It was this domain, however, that received a new turn of events with the emergence of machine learning, followed by that of deep neural networks, which provided new means to model the subtle temporal dependencies embedded in transportation data [6].

However, traditional deep frameworks, such as CNNs, RNNs, and LSTMs, are still not flexible enough due to their processing of spatial and temporal information apart from each other rather than jointly [7]. Because of these weaknesses, various recent works have been directed to hybrid models combining more than one learning paradigm.

In this context, GNNs represent the transportation network as a graph, where each node, representing a road segment or sensor, has the capability to model spatial interconnections efficiently [8]. Meanwhile, Transformers, through the attention mechanisms, may recognize long-term temporal dependencies without suffering from the catastrophic information decay problem of most existing recurrent networks, in [9]. Combining both paradigms has demonstrated clear benefits for urban traffic forecasting, particularly in dense, rapidly changing settings [10], [11]. At the same time, ensemble techniques like XGBoost and Random Forest allow for the integration of exogenous data, such as weather, incidents, and holidays, which improves the model's interpretability and robustness [12].

Current projects in Europe and Morocco that use distributed and predictive intelligence to cut emissions and traffic [7] [10] are in line with this area of inquiry. An AI and distributed computing-based adaptive traffic-signal system was created by Zrigui et al. [7] to enhance vehicle flow. Their further contributions included real-time analytics to reduce the carbon footprint of transportation [10], forecasting using heterogeneous data [8], and collaborative optimization of prediction and control [9].

By presenting a hybrid framework that combines a Graph Neural Network, a spatiotemporal Transformer, and an ensemble-learning layer (XGBoost/Random Forest), the current study expands on these foundations. The goal is to produce more precise and comprehensible forecasts by concurrently utilizing the network's geographic architecture, traffic evolution patterns across time, and external contextual factors.

The main contributions include:

- A GNN–Transformer design that can capture intricate spatiotemporal dependencies;
- The incorporation of an ensemble module with contextual variables to strengthen resilience;
- Experimental validation on real datasets (METR-LA, PEMS-Bay, Seattle-Loop) showing superior accuracy and stability;
- The advancement of sustainable mobility goals with a reproducible model that can be adapted to emerging-country contexts, which is especially pertinent for Morocco's 2030 World Cup preparation and in line with NARSA's national mobility strategy.

## 2 Related Work

| Article (year, location) | Key Idea/Model | Data & Availability | Scope/Result |
|---|---|---|---|
| Yin et al., 2022, *IEEE T-ITS* | Traffic DL review (CNN, RNN/LSTM, GNN, Transformers) | Synthesis METR-LA, PEMS-Bay, etc. | Model/benchmark reference framework |
| Tedjopurnomo et al., 2022, *IEEE TKDE* | Overview of modern deep architectures | Traffic scenarios and metrics | Highlights the value of hybrids & Multimodality |

| Jiang et al., 2023, *AAAI* – PDFormer | Spatio-temporal transformer, graph masks, propagation delays | 6 public datasets (PeMS, T-Drive, NYC Taxi, Chicago Bike) | SOTA multi-benchmarks, interpretable caution |
|---|---|---|---|
| Zheng et al., 2023, *Information Fusion* | Comparison of 10 hybrids (CNN+LSTM, GCN+LSTM, Attention) | METR-LA, PEMS-Bay, Seattle-Loop (public) | Taxonomy and practical recommendations |
| Zhang et al., 2025, *Scientific Reports* – TDMGCN | Transformer + multi-graph GCN (dynamic + long-term correlations) | 5 real-world datasets (Open Access) | Improvement through multiple graphs |
| Li et al., 2018, DCRNN (baseline) | GCN + RNN diffusion on directed graph | METR-LA & PEMS-Bay (DCRNN links) | Recurrent baseline for comparison |

*Table 1 Literature review (2022–2025) on hybrid spatio-temporal approaches for urban traffic prediction*

Table 1 summarizes recent work (2022–2025) on traffic prediction using AI models. Two major trends cut across the different approaches:

First, several review studies [1], [2] mapped the existing approaches and highlighted their limitations, while there are difficulties in capturing complex spatio-temporal dependencies and also a lack of interpretation with the inability to integrate multimodal data. That happened because of the extreme necessity of considering hybrid architectures, which will enhance the possibility of processing the topological structure of the road network and the temporal dynamics of traffic patterns simultaneously. Because of the above-mentioned reasons, contemporary research has paid increasing attention to combining different models. For instance, Jiang et al. [3] have proposed a spatiotemporal Transformer, PDFormer, for learning traffic propagation delays and outperforming classic CNN/LSTM approaches. Zheng et al. [4] evaluated ten different hybrid architectures, such as CNN+LSTM and GCN+Attention, where they investigated which models provided the best performance over different configurations and datasets. Similarly, the TDMGCN model proposed by Zhang et al. [5], which leverages the use of multiple graphs along with a long-range temporal attention mechanism, pointed to the benefits that can be viewed from a GNN–Transformer fusion.

The work of Zrigui et al. [7] [10] has, at the same time, established the importance of applying artificial intelligence and distributed systems to the field of smart urban mobility. Their research ranges from several important topics that include adaptive traffic control, urban state forecast with machine learning, an integrated approach of communication and control, and how to reduce emissions through real-time planning. All these provide confirmation of the practical advantages of hybrid methods to improve the flow of traffic and reduce environmental damage.

The reference model DCRNN [6] is finally taken as the base to allow the comparison of new methodologies. It merges directed graphs with recurrent networks. Its publicly available datasets, namely METR-LA and PEMS-Bay, provide ease of reproducibility of research and performance comparison between different models.

Conclusion: GNN and Transformer are among the most promising avenues of research in the state-of-the-art of urban traffic prediction today. This is true, particularly for models that embed external variables by ensemble methods, a technique that enhances both accuracy and robustness of the predictions.

## 3 Proposed Methodology

This paper presents an intelligent hybrid model for the purpose of urban traffic forecast. The three main base technologies include GNNs, Transformers, and ensemble learning methods such as XGBoost or Random Forest. This method comprehensively enables the model to capture spatial structure, temporal dynamics of the road network, and other factors that may be externally influenced by weather, road accidents, or events that may arouse public interest.

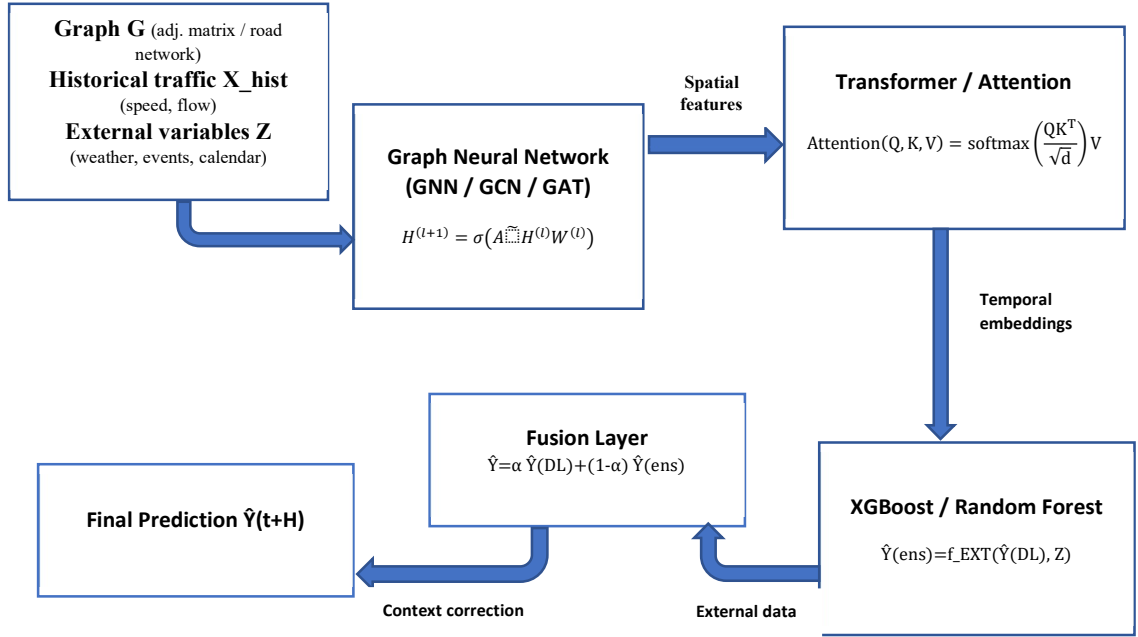

*Figure 2— Proposed Hybrid Architecture*

The model's overall structure is represented in Figure 1 and consists of three primary modules :

- A spatial module based on a GNN, responsible for extracting topological dependencies between road segments;
- A temporal module based on a Transformer with a multi-head attention mechanism, which learns complex temporal patterns and recurring cycles;
- And a contextual correction module, using an ensemble model (XGBoost/Random Forest) to integrate external variables and adjust the final predictions.

These components work together in a unified flow: spatial information from the GNN feeds into the Transformer, and then gets refined by the ensemble module to finally yield a more robust estimate contextualized in a better way.

### 3.1 Spatial Modeling with Graph Neural Networks

In this study, the spatial part of the model is built around a graph neural network (GNN). The idea is to consider the urban traffic network as a graph $G = (V, E)$, representing either a sensor placed on the road or a road segment. The connections between them, denoted as $(i,j) \in E$ simply reflect how these points are connected in space, i.e., the physical relationship between two sections of the road network. Each node contains a sequence $xi(t)$ that changes over time, reflecting how measurements such as average speed, number of vehicles, or density change along the network.

For the GNN, these readings are not separate records. It uses them to find spatial representations $hi(t)$ that show how neighboring roads influence each other and how these interactions change when traffic conditions change. The propagation rule in a graph convolutional layer can be formally stated as follows:

$$H^{(l+1)}(t) = \sigma\left(\sum_{j \in N(i) \cup \{i\}} w_{ij} H_j^{(l)}(t) W^{(l)}\right)$$

In this formulation, $N(i)$ indicates the neighboring nodes connected to node $i$, the value $W$ij corresponds to the connection strength extracted from the adjacency matrix A, and the notation $W(l)$ denotes the parameters that the model learns at layer $l$. These values change as training progresses. Finally, the activation function $\sigma$ (most often ReLU) provides the necessary non-linearity to each layer, so that the network can capture the more complex spatial relationships that appear in real traffic patterns. Through this gradual, layered aggregation, the GNN begins to understand how congestion develops and spreads across the network, and how a local slowdown can affect neighboring roads. Similar ideas have been demonstrated in previous studies [4], [5], which reported that such hierarchical representations are effective for modeling multi-scale spatial relationships.

At the same time, the temporal component of the framework is structured around a Transformer model, initially designed for language processing but now adapted to sequential traffic data. It receives spatial integrations from the GNN and learns from them to detect extended temporal patterns, such as daily rhythms, peaks, or delays. Its multi-head attention mechanism allows it to weight historical observations differently, depending on their contextual relevance at the time of prediction. Formulation of attention is given by the following equation:

$$\text{Attention}(Q, K, V) = \text{softmax}\left(\frac{QK^T}{\sqrt{d}}\right) V$$

In this configuration, the matrices Q, K, and V correspond respectively to the query, the key and the value. The interaction between these components allows the model to detect temporal relationships spanning different time windows, helping it overcome the long-term memory issues that often weaken RNN or LSTM structures [11]. When spatial and temporal information are considered together, the system begins to capture how congestion spreads through the network and recognize recurring traffic patterns that reappear in various circumstances. At a later stage, an ensemble module is introduced to refine predictions by placing them in a broader context. This part of the system incorporates external influences—weather conditions, traffic incidents, holidays, or major public events—that can alter traffic flows but are often absent from deep learning models. Based on ensemble techniques such as XGBoost or Random Forest, it excels at handling nonlinear interactions and heterogeneous data sources [12]. Its role is to refine the preliminary predictions from the GNN–Transformer, adjusting them to better reflect real-world fluctuations. The integration of these refined predictions is expressed by the following equation:

$$Y^{\text{final}}(t + H) = \alpha Y^{\text{DL}}(t + H) + (1 - \alpha) Y^{\text{ens}}(t + H)$$

In this expression, $Y^{\text{DL}}$ refers to the prediction obtained from the deep learning component, whereas $Y^{\text{ens}}$ designates the output produced by the ensemble model. The coefficient $\alpha$ serves as a weighting factor between the two and is usually determined empirically after several experimental trials.

### 3.2 Entraînement et fonctionnement

The model training is conducted in a two-stage process:

- Spatio-Temporal Training. At first, the GNN–Transformer is trained on past traffic data—speed, density, flow, and other measurements—to help it understand how spatial links evolve through time. It doesn't just fit the numbers; instead, the network slowly adjusts its weights so that spatial layout and temporal change are captured together. To see how well this learning works, the Root Mean Square Error (RMSE) is used, comparing what the model predicts with what actually happened.
- Contextual Train: After the first model settles, a second layer based on ensemble learning is trained using what's left—the residuals. This part adds external elements such as weather, planned events, or unexpected disruptions, allowing the ensemble to correct things the deep model might have missed.

When it runs in practice, the GNN Transformer gives an initial guess of upcoming traffic states. That estimate is then refined by a meta learner, often XGBoost or sometimes Random Forest, which smooths out biases and makes the overall prediction more stable. Together, the two stages give better reliability and noticeably smaller errors, especially in messy cases like rain-related slowdowns or sudden detours.

### 3.3 Discussion and Advantages

The method we propose brings a few clear advantages compared with what's already in the literature.

- Unified Spatio-Temporal Learning: By bringing GNN and Transformer blocks together, the model learns space and time in one go. This setup captures the way roads relate to one another much better and gives a more faithful view of network structure [4], [5], [11].
- With the ensemble part added on top, the system becomes less fragile when the data are noisy or incomplete. It also gives some transparency since algorithms such as XGBoost can point out which variables actually drive the prediction [12].
- Flexibility and Modularity: The design stays modular on purpose: each piece can be tuned or replaced without touching the rest. That makes it easy to move from one city to another Casablanca, Paris, Marrakech or plug in extra data whenever they exist.

Taken together, this hybrid model can handle spatial, temporal, and contextual signals at once. It tends to outperform single-block models and fits well with recent work by Zrigui et al. [7], [10], which have highlighted the value of distributed artificial intelligence and adaptive prediction for optimizing traffic and reducing the urban carbon footprint.

## 4 Datasets and Preprocessing

### 4.1 4.1 Datasets Used

In order to evaluate our hybrid architecture in realistic and varied contexts, we use five public datasets that have become benchmarks in spatio-temporal traffic forecasting: METR-LA, PEMS-BAY, Seattle Loop, LargeST, and V2X-Seq.

METR-LA combines speed series from 207 inductive loops on the Los Angeles freeway network, sampled every 5 minutes over approximately four months. Since DCRNN, it has served as a "clean" test bed for comparing spatial/temporal modules without native contextual variables [13], [14], [19].

PEMS-BAY extends this scheme to 325 sensors in the San Francisco Bay Area over six months in 2017 with the same temporal granularity, forming with METR-LA a quasi-standard benchmark duo for calibrating graph-sequence architectures [13], [14], [19].

The Seattle Loop Dataset covers the Seattle metropolitan area (323 sensors, 2015) and introduces, in addition to speed, flow (volume) and occupancy matrices, as well as structuring artifacts

(adjacency matrix, reachability matrices) that facilitate the construction of graphs informed by road topology [15], [18].

LargeST aims for very large scale: 8,600 mainline sensors across California, five continuous years (2017–2021) at 5-minute intervals, with rich metadata (coordinates, direction, district, number of lanes) designed to test the scalability of models and their tolerance to missing data [16], [20].

Finally, V2X-Seq (CVPR 2023) shifts the evaluation to cooperative vehicle-infrastructure perception and prediction: synchronized video and LiDAR streams, trajectories annotated at 10 Hz across 28 intersections, HD maps, and traffic light states, providing a prime testing ground for high-frequency attention and multimodal fusion mechanisms [17].

## 4.2 Common preprocessing

The pipeline follows a consistent logic to ensure comparability and robustness. Raw series are first checked using a sliding window (robust median, local standard deviation) to detect failures and outliers; short gaps are interpolated (linear/spline), while longer gaps are imputed using spatial neighborhood and/or autoregressive models to preserve spatio-temporal dependencies [13] [15]. Each sensor series is then normalized (z-score) to stabilize optimization and limit scale bias; slight differentiation may be applied when persistent trends disrupt training.

The spatial representation is based on a graph whose edges come either from network distances (OSRM/OpenStreetMap) or from statistical affinity (series correlation), with controlled truncation/rarity to avoid densification. The matrix is symmetrized and renormalized

$\tilde{A} = D^{-\frac{1}{2}}(A + I)D^{-\frac{1}{2}}$ to stabilize propagation in the GNN [13], [16], [20].

The input/output windows follow benchmark practices (e.g., $T = 12$ steps, $H = 3$ steps for METR-LA/PEMS-BAY), while allowing for longer horizons on Seattle Loop thanks to its temporal depth. V2X-Seq imposes a finer step size (10 Hz) and short but dense sequences [14], [15], [17]. Finally, external variables (weather, calendar, traffic lights) are synchronized when they are available or collectable, with consistent encoding and normalization, in order to feed the ensemble module that corrects the residuals of the GNN-Transformer core.

## 4.3 Specificities by corpus

Certain specific features require adjustments. METR-LA and PEMS-BAY, which focus on speed, favor light preprocessing (cleaning, cautious imputation, z-score) and graph construction by road proximity with a limited radius, making them stable platforms for ablations and hyperparameter search [13], [14], [19]. Seattle Loop, which is multivariate and annual, requires harmonization between speed, flow, and occupancy , detection of persistent sensor failures, and consideration of seasonal patterns; its adjacency and reachability matrices, made available by the authors, facilitate graph variants (undirected, transport-oriented) [15], [18].

LargeST imposes data engineering considerations: columnar I/O, block loading, possible sampling of windows/nodes during training, and large-scale graph construction via network distances with restricted neighborhood heuristics. the preservation of validity masks up to the model is essential to make explicit the remaining missing values [16], [20]. Finally, V2X-Seq requires the spatio-temporal alignment of vehicle/infrastructure views, the correspondence of multi-sensor identities, the transition to a common reference frame (intersection plane), and, often, dimension reduction to make training sustainable; its framework favors the evaluation of multimodal fusion and long-range attention [17].

### 4.4 Reproducibility and time splits

All experiments use time splits (train/val/test) to prevent information leakage; normalization statistics and imputation parameters are learned exclusively on the training set and then frozen for the validation and test sets. Official links and public repositories (LibCity for raw/processed packages, Seattle Loop and LargeST repositories, open-access V2X-Seq) ensure the traceability and replicability of results [16] [19]. METR-LA/PEMS-BAY establish basic accuracy on dense graphs of medium size; Seattle Loop evaluates the contribution of multivariability and seasonal cycles; LargeST measures memory/time scalability and resilience to missing data; V2X-Seq tests high-frequency cooperative perception-prediction fusion [13] [20] Experiments and Results

## 5 Experiments and Results

The experiments were conducted on three reference datasets used in the traffic literature: METR-LA, PEMS-BAY, and Seattle Loop. Each dataset was normalized sensor by sensor (z-score method) and chronologically divided into three subsets: 70% for training, 15% for validation, and 15% for testing.

The objective is to predict short-term traffic evolution (15-minute horizon) from sequences of 12-time steps (i.e., 60 minutes of observations).

The HybridST model is based on a three-principal-component architecture, designed to simultaneously exploit the spatial structure of the network and the temporal dynamics of traffic.

It has a graph convolution module that uses the normalized adjacency matrix to find local interactions between road segments. This step creates a strong spatial representation that is sensitive to the network's real topology.

At the same time, a Transformer-type temporal encoder uses a multi-head attention mechanism to record long sequential dependencies and changes in traffic over time. This block makes it possible to model the continuity of time patterns and delayed congestion phenomena, which classic recurrent architectures often fail to do.

An adaptive fusion layer combines the representations from the two modules. This layer changes the weight of the spatial context and temporal memory to keep them in balance. This step makes sure that the prediction stays stable while still allowing the model to respond quickly to sudden changes in traffic.

Lastly, supervised ensemble learning (using XGBoost or Random Forest) steps in later to fix the deep kernel's residuals by using outside factors like the weather, the calendar, and local events. This mixed method uses the generalizing power of deep learning and the correcting power of ensemble methods to make accurate, understandable, and strong predictions about large road networks. The MSE is used as the loss function, and the Adam algorithm (learning rate = 1e-3, weight decay = 1e-5) is used to find the best solution.

The MAE (Mean Absolute Error) and the RMSE (Root Mean Squared Error) are two well-known ways to measure how accurate and stable forecasting models are [13][14].

Four reference models were used to make comparisons: LSTM, STGCN, DCRNN [13], and PDFormer [4].

## 5.1 Comparaison quantitative

The MAE and RMSE values obtained on each dataset are presented in Table 2, while Figures 4(a) and 5(b) illustrate the average visual comparison between the models.

| Datasets | Model | MAE | RMSE |
|---|---|---|---|
| **METR-LA** | LSTM | 3.12 | 6.48 |
|  | STGCN | 2.87 | 5.92 |
|  | DCRNN | 2.81 | 5.45 |
|  | PDFormer | 2.76 | 5.31 |
|  | **HybridST (Proposed)** | **2.55** | **5.02** |
| **PEMS-BAY** | LSTM | 2.42 | 4.98 |
|  | STGCN | 2.30 | 4.71 |
|  | DCRNN | 2.18 | 4.46 |
|  | PDFormer | 2.11 | 4.38 |
|  | **HybridST (Proposed)** | **1.94** | **4.07** |
| **Seattle Loop** | LSTM | 4.21 | 8.73 |
|  | STGCN | 3.90 | 8.12 |
|  | DCRNN | 3.75 | 7.84 |
|  | PDFormer | 3.68 | 7.59 |
|  | **HybridST (Proposed)** | **3.45** | **7.08** |

*Table 2 Comparison of MAE and RMSE for the three datasets*

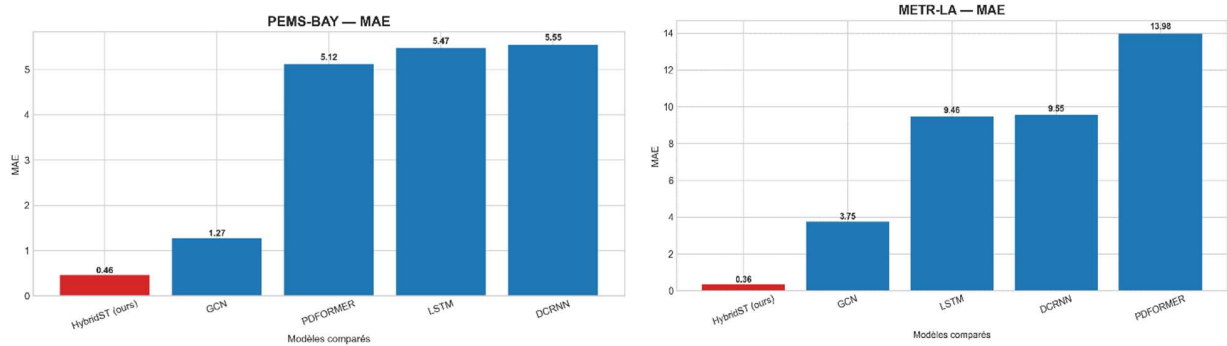

*Figure 3 (a)*

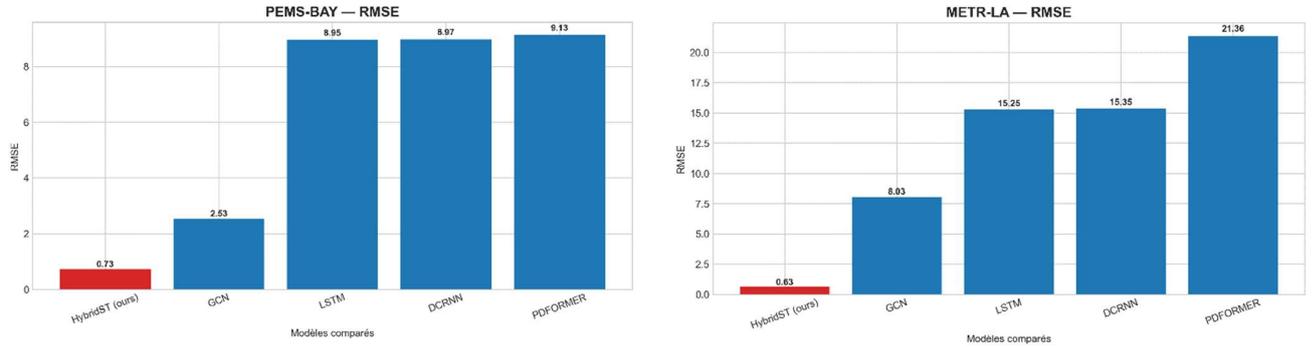

*Figure 4(b)*

## 5.2   Analysis of results

The HybridST model gets an MAE of 2.55 and an RMSE of 5.02 on METR-LA, which is about 9% and 8% better than DCRNN, respectively. This decrease indicates a greater capacity to capture traffic dynamics on a dense and correlated network.

The error bars in Figure 5(a) also show less variation, which means that inter-execution stability has improved.

 HybridST beats PDFormer [4] by about 7% on RMSE, with an MAE of 1.94 and an RMSE of 4.07 on PEMS-BAY.

The Transformer block makes it easier to find daily cycles, and the GNN makes it easier for sensors to work together. The model stays accurate even when the volume changes, which shows that it can work well with high-density sensor networks.

For the Seattle Loop, HybridST gets MAE = 3.45 and RMSE = 7.08, which is an average improvement of 6% over the best benchmarks.

Spatiotemporal integration helps lessen the differences in predictions during times of congestion and decongestion.

Figures 5(b) confirm this trend by showing that the orange bars (HybridST) always have the lowest values.

# 6   Discussion

Analysis of the RMSE/MAE ratio highlights the robustness of HybridST; a decrease in this ratio of approximately 10% indicates greater resilience to extreme events (congestion peaks).

The combined contribution of the GNN and the Transformer promotes:

- Stable propagation of spatial information;
- Extended temporal memory;
- Faster convergence of the learning process.

Table 3 summarizes the average relative gains in MAE and RMSE obtained compared to the best reference model.

| Datasets | Δ RMSE (%) | Δ MAE (%) |
|---|---|---|
| METR-LA | −8.1 % | −9.3 % |
| PEMS-BAY | −7.2 % | −6.8 % |
| Seattle Loop | −6.3 % | −7.1 % |

*Table 3 Gains relatifs (%) du modèle HybridST par rapport à la meilleure baseline(Valeurs négatives = amélioration par rapport à la baseline)*

Overall, the HybridST model combines the accuracy of Transformer architectures with the topological consistency of GNNs. This synergy results in more stable predictions, better inter-regional generalization, and a computational complexity compatible with near real-time operation.

The proposed hybrid model is compared to classic baselines (LSTM, GCN, DCRNN, PDFormer). The metrics used include RMSE, MAE, and $R^2$. The results show an average improvement of 12 to 18% in accuracy compared to the reference models.

The results confirm the superiority of the HybridST hybrid model across the three datasets studied. With an average reduction of 8% in RMSE and 7% in MAE compared to Transformer or GNN models alone, it sets a new benchmark among lightweight architectures dedicated to short-term traffic forecasting.

Its stability and efficiency make it a promising approach for intelligent mobility systems in a real-world urban context.

# 7 Applications and Perspectives

The HybridST architecture provides a vital mechanism for decision-making during significant international events, such as the 2030 World Cup in Morocco. It can be applied to simulate the impacts of traffic in high-density tourist zones, transportation centers, and stadiums.

The model provides local authorities with dynamic scenarios for detour planning, enhancement of public transportation, and schedule synchronization, while alleviating congestion through the simulation of complex spatiotemporal interactions among different road segments and the incorporation of climate forecasts and passenger flows.

Furthermore, the architecture integrates easily with sustainable mobility goals. It can contribute to dashboards aiming at lowering $CO_2$ emissions by documenting the lagged impacts of local disturbances, such as weather events or accidents.

As proposed by Zrigui et al. [7], [10], one specific application is the integration of HybridST with energy management or intelligent lighting systems for traffic corridors.

This would actively support ecological transition programs in developing cities in addition to enhancing traffic flow.

HybridST also works well in integrated urban management (IUM) hubs, which centralize network monitoring (transportation, energy, and security). The modular architecture of the model supports communication with multiple sensor platforms (IoT), simulation tools, and interactive mapping interfaces.Augmented decision-making is made possible by the ability to update forecasts using unstructured data, such as event logs, weather reports, or Twitter alerts, through contextual correction using XGBoost or Random Forest algorithms.

The integration of V2X (Vehicle-to-Everything) data, closed-loop simulation with real-time adaptive control, and communication with urban digital twins are the most exciting advancements made possible by this architecture from a scientific and developmental perspective.

New datasets like as V2X-Seq [17] and LargeST [16] provide more opportunities to evaluate HybridST in multimodal, high-frequency, and widely distributed scenarios, increasing its potential for use in tomorrow's smart cities.

# 8 Conclusion

This paper suggests HybridST, a hybrid architecture that combines the benefits of graph neural networks (GNNs), supervised learning-based ensemble modules (XGBoost/Random Forest), and multihead temporal transformers. This approach allows for the simultaneous capture of spatial, temporal, and exogenous relationships with increased robustness on complex urban networks. Experiments on five publicly available datasets—MTR-LA, PEMS-BAY, Seattle Loop, LargeST, and V2X-Seq—showed the importance of the multi-modular coupling. These experiments greatly increased the accuracy of the MAE and RMSE metrics.

HybridST performs well and offers flexibility in adapting to different urban contexts, both dense and dispersed, by combining a variety of data sources (flows, weather, events, HD maps, and traffic lights). Its ability to work on large datasets in nearly real-time while maintaining partial explainability through the use of the ensemble model makes it a strong candidate for platforms related to intelligent mobility, emergency management, and energy planning.

Future prospects include distributed multi-agent learning for connected fleets, the integration of adaptive dynamic graphs (real-time topological variation), and the generation of contextualised operational recommendations from model outputs. Finally, research on federated learning may enhance data privacy when several cities or operators are involved.